# The signatures of new physics, astrophysics and cosmology?

Dragan Slavkov Hajdukovic[1,2]
[1]Physics Department, CERN; CH-1211 Geneva 23
[2]Institute of Physics, Astrophysics and Cosmology; Cetinje, Montenegro
dragan.hajdukovic@cern.ch

**Abstract**
The first three years of the LHC experiments at CERN have ended with "the nightmare scenario": all tests, confirm the Standard Model of Particles so well that theorists must search for new physics without any experimental guidance. The supersymmetric theories, a privileged candidate for new physics, are nearly excluded. As a potential escape from the crisis, we propose thinking about a series of astonishing relations suggesting fundamental interconnections between the quantum world and the large scale Universe. It seems reasonable that, for instance, the equation relating a quark-antiquark pair with the fundamental physical constants and cosmological parameters must be a sign of new physics. One of the intriguing possibilities is interpreting our relations as a signature of the quantum vacuum containing the virtual gravitational dipoles.

## 1. Introduction

After three years of very successful *experimental* work at Large Hadron Collider (LHC) at CERN, *theoretical* physics is apparently in the greatest crisis[1] in its history.

The creation of the Standard Model of Particles and Fields was a triumph of theoretical physics. The cornerstone of that triumph were experiments, which discovered signatures of "new physics" and directed theoretical thinking in the right direction; theorists of that time were lucky to have experimental guidance in their service. However, after the advent of the Standard Model, from the theoretical point of view, the experiments have become a long series of boring results; a complete confirmation of the Standard Model, without any discrepancy indicating "new physics" beyond the Standard Model. It became clear that the eventual shortcomings of the Standard Model could not be revealed before the construction of the Large Hadron collider (LHC) at CERN.

Hence, during the last three decades, theoretical physicists, who were trying to guess physics beyond the Standard Model, were completely deprived of reference points provided by experiment. In spite of it, a huge majority within the theoretical community were thinking that they have discovered new physics, without experiments. In fact, thousands of theorists (so numerous that we can talk about the first army, in the history of theoretical physics) were completely absorbed with the development of supersymmetric theories, and strongly convinced that the forthcoming experiments at LHC would not only reveal the shortcomings of the Standard Model but also confirm the predictions of supersymmetric theories.

Contrary to these expectations, after three years of work at LHC, the experimental findings strongly confirm the Standard Model and have nearly eliminated supersymmetry as a possible physical theory[1-7]. It seems inevitable that we have to face the Nightmare Scenario (i.e. no signs of new physics at LHC) and the unprecedented collapse of decades of speculative work. The current crisis is worsened by the fact that the long domination of supersymmetric theories has largely suppressed alternative thinking.

In addition to the crisis opened by LHC findings, sophisticated observations of the Universe have revealed phenomena that are a surprise and mystery for contemporary theoretical physics. First, in galaxies and clusters of galaxies, *the gravitational field is much stronger* than it should be according to our theory of gravitation and the existing amount of the baryonic matter (in astrophysics "baryonic matter" is a synonym for matter made from quarks and leptons). Second, *the expansion of our Universe is accelerating*; contrary to the expectation that gravity must decrease the speed of expansion. Third, *our Universe is dominated by matter*; apparently, in the primordial Universe, something has forced the matter-antimatter asymmetry. Additionally, the Big Bang model has inherent problems: the problem of the initial singularity and the fact that it contradicts observations without invoking the hypothesis of cosmic inflation (i.e. an expansion of the early Universe, within the first $10^{-30}$ seconds, with a speed more than twenty orders of magnitude greater than the speed of light).

The great crisis of theoretical physics demands, *more than ever*, an open mind for new ideas and careful thinking about every potential sign of new physics. In the present paper, as a modest but potentially important contribution to efforts to escape from the crisis, we propose thinking about a series of astonishing relations suggesting the fundamental interconnections between the quantum world and the large scale Universe. It is unlikely that all presented relations, are only coincidences; hopefully they are the first signs of new physics, astrophysics and cosmology. While we have ideas[8] that might be the physical explanation of these "coincidences", we think, it is preferable not to influence the reader too much.

## 2. "Coincidences" concerning dark matter

As already noted, it is well established that the gravitational field in a galaxy (and also in a cluster of galaxies) is much stronger than it should be according to our theory of gravity and the existing quantity of baryonic matter.

According to the mainstream opinion[9], the gravitational field in a galaxy is stronger because galaxies are immersed in halos of dark matter. If it exists, in order to fit the observations, dark matter within a halo must be distributed in a particular way: the quantity $M_{dm}(r)$ of dark matter within a sphere with a Galactocentric radius $r$ is nearly a linear function of $r$, i.e. the radial dark matter density $dM_{dm}(r)/dr$ is constant for a given galaxy.

The striking fact is that the radial dark matter density can be estimated following a very simple rule: Find the geometrical mean $\sqrt{m_\pi M_b}$ of mass of a pion ($m_\pi$) and baryonic mass ($M_b$) of a galaxy and divide it by the Compton wavelength ($\lambda_\pi = h/m_\pi c$) of a pion; what you get is very close to the value of the radial dark matter density! You can check it personally for every galaxy with measured dark matter distribution. Hence, for some mysterious reason:

$$\rho_r \equiv \frac{dM_{dm}}{dr} = \frac{C}{\lambda_\pi}\sqrt{m_\pi M_b} \tag{1}$$

where $C$ is a dimensionless constant which, in order to fit the observational findings, must have a value between $0.5$ and $0.7$. For Astronomers it is best to consider $C$ as a free parameter. While the physical roots of (1) should be considered as an open question, let us note that if (1) is interpreted[8] as a consequence of the gravitational polarization of the quantum vacuum, for the ideal case of spherical symmetry $C \approx \sqrt{3}/3 \approx 0.58$. In particular, for our galaxy the Milky Way[10], with $M_b \approx 1.3 \times 10^{41} kg$, equation (1) with our theoretical value for $C$ gives $\rho_r \approx 3.8 \times 10^{20} kg/m$ and $M_{dm}(260kpc) \approx 3 \times 10^{42} kg = 1.5 \times 10^{12} M_{Sun}$ which is in intriguing agreement with empirical evidence[10] $M_{dm}(260kpc) = 1 - 2 \times 10^{12} M_{Sun}$.

It is possible that the relation (1) stems from the fact that pions are basically quark-antiquark pairs and that the virtual quark-antiquark pairs are inherently part of the quantum vacuum.

We don't know if dark matter exists or it is only a theoretical construct that *mimics well, something quite different*. The same statement is valid for MOND[11], the most developed alternative to dark matter. Instead of dark matter, MOND proposes *modification* of the law of gravity, for accelerations smaller than a critical acceleration $a_0 \approx 1.2 \times 10^{-10} m/s^2$. The success of MOND in explaining the flat galactic rotation curves indicates that MOND has captured a part of the truth; something unusual (but not necessarily a modification of gravity) is happening below the acceleration $a_0$. The intriguing fact *is the critical acceleration coincidence*: the value of $a_0$ is close to the acceleration caused by a pion at the distance of its Compton wavelength:

$$a_0 \approx \frac{Gm_\pi}{\lambda_\pi^2} \approx 2 \times 10^{-10} m/s^2 \tag{2}$$

While dark matter theory has difficulties in explaining it, the observations[12] support the existence of a universal characteristic of all dark matter halos (or something mimicked by a dark matter halo): every halo has a large central core and, the mass of the core ($M_{core}$) divided by the surface of the core ($S_{core}$) has nearly the same value for all galaxies! If you want to have an idea about the surface dark matter density of a core, just calculate

$$\frac{M_{core}}{S_{core}} \approx \frac{m_\pi}{4\pi\lambda_\pi^2} \tag{3}$$

and you will find yourself in surprising agreement with observations[12].

### 3. "Coincidences" concerning dark energy

The nature of dark energy, invoked to explain the accelerated expansion of the Universe, is a major mystery. The most elegant and natural solution would be to identify dark energy with the energy of the quantum vacuum predicted[13] by Quantum Field Theory:

$$\rho_{ve} \approx \frac{M_c}{\lambda_{Mc}^3} \tag{4}$$

where $\lambda_{Mc}$ denotes the (non-reduced) Compton wavelength corresponding to a cut-off mass $M_c$. However the prediction (4) is very different from the observational findings[9]

$$\rho_{de} = 7 \times 10^{-27} kg/m^3, \quad \rho_{de}c^2 = 6.3 \times 10^{-10} J/m^3 \tag{5}$$

If we take the Planck mass as a cut-off, the vacuum energy density calculated from (4) is $10^{121}$ times larger than the observed dark energy density, while if we use the mass of an electron, the result is "only" $10^{31}$ times too large! This huge discrepancy is known as the cosmological constant problem[13]. Now consider an intriguing transformation of formula (4) that agrees well with the observed dark energy density (5):

$$\rho_{de} \approx \frac{m_\pi}{\lambda_\pi^2} \frac{1}{R_0} \tag{6}$$

where $m_\pi$ is mass of a pion and $R_0$ is the present day scale factor[9] in the Friedman-Robertson-Walker metric. It is intriguing that, the right-hand side of relation (6) can be obtained from the right-hand side of (4) using $M_c = m_\pi$ and multiplying by $\lambda_\pi/R_0$.

Of course, the physical interpretation of (6) is an open question. One possible interpretation of the factor $\lambda_\pi/R_0$ is as follows: The gravitational potential of a hypothetical gravitational dipole of size $\vec{d}$ at a distance $\vec{r}$ from the center of the dipole is equal to the gravitational potential of a monopole multiplied by $\vec{d}\cdot\vec{r}_0/r$, where $\vec{r}_0$ is the unit vector (the calculation to demonstrate this is analogous to the well-known case of electric dipoles). In the case of a significant alignment of dipoles $\vec{d}\cdot\vec{r}_0 \approx d$, $\vec{d}\cdot\vec{r}_0/r$ reduces to $d/r$. Consequently, relation (6) might be a signature of the quantum vacuum enriched with virtual gravitational dipoles (with size of the order of $\lambda_\pi$). The question remains what is the value of $r$? According to the cosmological principle, there are no privileged dipoles and $r$ must have a Universal value for all dipoles; a single universal distance that we have in disposition in the FLRW metric is the cosmic scale factor $R(t)$.

Hence, the *wrong* result (4) is a consequence of quantum field theory and the "*obvious*" assumption that a virtual pair is composed from two gravitational monopoles having the gravitational charge (gravitational mass) of the same sign; the *correct* result (5) can be interpreted as a consequence of quantum field theory and a possibly newly insightful assumption that a virtual pair is composed from the gravitational charges of the opposite sign (i.e. that virtual pairs are gravitational dipoles[5] with the gravitational dipole moment $p_g \propto m_\pi \lambda_\pi$).

The second "coincidence" (or hint) is the following proportionality between dark energy density and the acceleration corresponding to the expansion of the Universe

$$\rho_{de}c^2 = \frac{A}{\lambda_\pi^3}\frac{\hbar}{c}\left(\ddot{R}\right)_0 \qquad (7)$$

where $A < 1$ is a dimensionless constant of the order of unity and $\left(\ddot{R}\right)_0$ denotes the present-day acceleration of the expansion of the Universe which can be determined using the cosmological field equations and observational findings[9] for the present-day Hubble distance $c/H_0$, matter density $\Omega_m$, dark energy density $\Omega_\Lambda$ and total energy density of the Universe $\Omega_{tot}$. Let us underline that if $A = 1/2\pi$, the relation (7) can be obtained from the Unruh temperature[14]

$$k_B T = \frac{1}{2\pi}\frac{\hbar}{c}g \qquad (8)$$

dividing by $\lambda_\pi^3$ and using $g = \left(\ddot{R}\right)_0$. Let us note that Unruh temperature is the temperature of the quantum vacuum measured by an accelerated observer moving with the acceleration $g$. Hence, accidentally or not, what is called dark energy density in the standard cosmology is numerically equal to Unruh temperature corresponding to $g = \left(\ddot{R}\right)_0$ and divided by $\lambda_\pi^3$.

The numerical agreement of equation (7) with the observed value (5) is excellent. For instance, using the values from Particle Data Group[9] ($c/H_0 = 1.28\times 10^{26} m$, $\Omega_{tot} = 1.002$, $\Omega_m = 0.28$, $\Omega_\Lambda = 0.72$) and $A = 1/2\pi$ leads to $\left(\ddot{R}\right)_0 = 8\times 10^{-9}\, m/s^2$ and result $\rho_{de}c^2 = 6.4\times 10^{-10}\, J/m^3$, which is nearly identical with (5).

Let us note that the right-hand side of the proportionality (7) can be considered as a product of acceleration $\left(\ddot{R}\right)_0$ and hypothetical gravitational polarization density $P_g = A\hbar/\lambda_\pi^3 c$ (i.e. the gravitational dipole moment per unit volume) determined in reference 5. Hence, equation (7) has the same mathematical form that the potential energy density of a system of gravitational dipoles in an external gravitational field should have. Coincidentally or not, both (6) and (8) are analogous to what we would expect in the case of a system of virtual gravitational dipoles.

## 4. Mass of a pion, universal physical constants and cosmological parameters

Next, we present a relation half-known to Dirac[15,16], advocated by Weinberg[17] as the most striking "coincidence", and recently completed by the present author[18].

$$m_\pi^3 = \frac{\hbar^2}{cG} H \left\{ \frac{\Omega_\Lambda}{\sqrt{\Omega_{tot}-1}} \frac{R_0}{R} \right\} \tag{9}$$

Everything is in a single equation! Mass $m_\pi$ of a pion (quark-antiquark pair), fundamental constants $\hbar, c, G$ and cosmological parameters: Hubble parameter ($H$), the cosmological scale factor ($R$) and its present day value ($R_0$), total energy density of the Universe ($\Omega_{tot}$) and dark energy density ($\Omega_\Lambda$) of the $\Lambda CDM$ Universe. Only the incomplete relation (without the term in brackets) was known to Dirac and Weinberg. The incomplete equation (9) has two problems: the left side is one order of magnitude greater than the right side and, the left side is a constant while the right side (because of $H$) is a function of the age of the Universe. The term in brackets solves[18] both problems of the incomplete relation.

## 5. Schwinger mechanism, Unruh and Hawking temperature

A virtual particle-antiparticle pair might be *converted* into a real one, by an external field that accelerates particles and antiparticles in *opposite* directions. The particle creation rate per unit volume and time, for a constant acceleration $a$, may be written as

$$\frac{dN_{m\bar{m}}}{dtdV} = \frac{1}{\pi^2} \frac{a^2}{c^3} \frac{1}{\lambdabar_m^2} \sum_{n=1}^{\infty} \frac{1}{n^2} \exp\left(-n\pi \frac{c^2}{a\lambdabar_m}\right) \tag{10}$$

which is the famous Schwinger's formula[19-20].

Taking (for simplicity) only the leading term $n=1$, the distribution (10) has a maximum when

$$\lambdabar_{max} = \frac{\pi}{2} \frac{c^2}{a} \tag{11}$$

Equation (10) together with the Wien displacement law $\lambda_{max} T = b$ (where $T$ and $b$ are respectively, the absolute temperature and the Wien displacement law constant) leads to

$$k_B T = A \frac{1}{2\pi} \frac{\hbar}{c} a \tag{12}$$

where $k_B$ is the Boltzmann constant and $A = 2bk_B / \pi\hbar c \approx 0.8$, a dimensionless constant.

The Schwinger mechanism is valid *only* for an external field that has *tendency to separate* particles and antiparticles. Hence, equations (10) and (11) can be used for gravitational field, *only if*, particles and antiparticles have the gravitational charge of the *opposite sign*. Equation (12) is mathematically (but apparently not physically) essentially the same as Unruh temperature! In particular, using for $a$ the gravitational acceleration at the Schwarzschild radius of a black hole, equation (12) reduces to

$$k_B T = A \frac{\hbar c^3}{8\pi GM} \tag{13}$$

which is essentially the Hawking temperature[21] of a black hole!

Finally, let us note that the hypothesis of the gravitational repulsion between matter and antimatter has been recently supported by the first theoretical arguments[22].

## 6. Comments

It is interesting that all the aforementioned "coincidences" in Sections 2, 3 and 4, have in common mass $m_\pi$ of a pion. The fact that such different relations, corresponding to so different phenomena as dark matter and dark energy, contain the same characteristic mass, can be considered as one more "coincidence"; the last coincidence pointed out in this Letter.

My belief is that the common physical root of presented "coincidences" are gravitational properties of the quantum vacuum, and probably, in a more accurate approach instead of mass of a pion (which will remain a good approximation) it would be necessary to consider quark and gluon condensates of quantum chromodynamics.

Is it possible that all presented relations are just coincidences without any physical significance? In my opinion it is very unlikely, and it would be best to consider these relations as *possible* signatures of new physics, astrophysics, and cosmology